\documentclass[aps,11pt,letterpaper,nofootinbib,showpacs,amsmath,amssymb,pdftex11pt]{revtex4}%
\usepackage{latexsym}%
\usepackage[dvips,dvipdfmx,hypertex]{hyperref}%

\def\diag{\text{diag}}
\def\tanh{{\text{tanh}}}

\DeclareMathAlphabet{\mathpzc}{OT1}{pzc}{m}{it}

\newcommand{\beq}{\begin{equation}}
\newcommand{\beqn}{\begin{equation}\nonumber}
\newcommand{\eeq}{\end{equation}}
\newcommand{\bea}{\begin{eqnarray}}
\newcommand{\bean}{\begin{eqnarray}\nonumber}
\newcommand{\eea}{\end{eqnarray}}

\begin{document}

\title{Quantum dust collapse in 2+1 dimension}
\author{Souvik Sarkar\footnote{\tt sarkarsi@mail.uc.edu}}
\author{Cenalo Vaz\footnote{\tt Cenalo.Vaz@uc.edu}}
\author{L.C.R. Wijewardhana\footnote{\tt Rohana.Wijewardhana@uc.edu}}
\affiliation{Department of Physics, University of Cincinnati, Cincinnati, OH 45221-0011.}

\begin{abstract}
In this paper we will examine the consequence of a canonical theory of quantum dust collapse in 2+1 dimensions. 
The solution of the WDW equation describing the collapse indicates that collapsing shells outside the apparent 
horizon are accompanied by outgoing shells within the apparent horizon during their collapse phase and stop collapsing 
once they reach the apparent horizon. Taking this picture of quantum collapse seriously, we determine a static 
solution with energy density corresponding to a dust ball whose collapse has terminated at the apparent horizon. 
We show that the boundary radius of the ball is larger than the BTZ radius confirming that no event 
horizon is formed. The ball is sustained by radial pressure which we determine and which we attribute to the 
Unruh radiation within it.

\end{abstract}
\maketitle

\section{Introduction}
In the mid-70's Bekenstein \cite{bek72} and Hawking \cite{haw75} studied the behavior of quantum fields in the neighborhood
of blackholes and argued that the latter would evaporate thermally by quantum effects so that quantum information would be 
lost during the evaporation process. This is the famous 'Information Paradox'. The fundamental postulates of Quantum 
Mechanics say that all the information about a quantum system is encoded in its wavefunction until the latter collapses. 
The evolution of the wavefunction is determined by an unitary operator and unitarity implies information is conserved in 
the quantum sense. However, if the system entering a blackhole is in a pure state, the transformation into the mixed state 
that describes the Hawking radiation would destroy information about the original quantum state. This leads to a breakdown of 
unitarity in quantum mechanics whenever an event horizon is present. From the 'No-hair theorem' it is expected that Hawking 
radiation is independent of the form of matter entering the blackhole. Therefore, this paradox is generic.

According to Quantum Field Theory (QFT) in curved spacetime, a single emission of Hawking radiation involves two mutually 
entangled states where an outgoing particle escapes as Hawking radiation and the infalling one is swallowed by the blackhole. 
Therefore the exterior is entangled with the interior. For an observer with access only to the exterior, the outgoing particle 
is in a mixed state and since the quantum numbers of the particle inside the blackhole can never escape, there will only be 
an exterior mixed state if the black hole evaporates completely. To resolve this paradox, several proposals have been made.

The most straightforward resolution to this paradox would be to assume that the evaporation process leaves behind a 
remnant by some, as yet unknown, mechanism. This is difficult to imagine because it requires that a relatively small object 
would possess a large degeneracy while remaining stable. A second option is to assume that Hawking radiation is in fact 
pure.  As proposed by Susskind et al \cite{sus}, this can be achieved if one requires that information is both 
emitted at the horizon and passes through it so that an observer outside would see it as the Hawking radiation and an 
observer who falls into it would see it inside, but no single observer would be able to confirm both pictures (so as to avoid 
cloning). Thought experiments \cite{thex} that support Blackhole Complementarity employ three assumptions: a) Hawking 
radiation is unitary b) Effective Field Theory (QFT in curved spacetime) is valid outside the event horizon and c) the 
Equivalence Principle holds. However, in 2012, Almehri, Marolf, Polchinsky and Sully (AMPS) \cite{amps12} argued that the 
three assumptions above, taken together, are logically inconsistent: if postulates a) and b) are assumed, then a Bogoliubov 
transformation from the frame of the distant, static observer for whom the quantum field is in a pure state to that of a 
freely falling observer indicates that the latter will see thermal radiation as she crosses the horizon. This violates the 
equivalence principle and leads to the AMPS ``firewall'' at the horizon. 

However Hawking \cite{haw12} proposed an alternative solution by suggesting the possibility that the collapse doesn't form 
an event horizon, rather matter stops collapsing once it reaches the apparent horizon. No singularity and event horizon will 
form and in the absence of an event horizon, the entire discussion of information loss becomes irrelevant.

We attempt to realize Hawking's proposal in a quantized model of dust collapse in $2+1$-dimensions with a negative cosmological 
constant. We do so because 2+1 dimensional dust collapse is already quite well understood on the classical and semiclassical levels
and the BTZ black hole is well understood on the quantum level via the AdS/CFT correspondence. Thus it may be possible to compare 
the degrees of freedom of the two approaches at a future date. Our approach here will be to exploit an exact canonical quantization 
of the Lema\^itre-Tolman-Bondi (LTB) family of solutions \cite{ltb} and examine the implications 
of the functional solutions to the Wheeler-DeWitt (WDW) equation \cite{wdw}. Two kinds of solutions are obtained. In one, 
matter coalesces on the apparent horizon from the interior and the exterior. In the second, matter moves away from the apparent 
horizon on both sides of it. In the first solution, the exterior infalling waves represent collapsing shells of dust, which 
are necessarily accompanied by interior outgoing waves representing the Unruh radiation. In the second solution continued 
collapse of the dust shells to a central singularity is accompanied by exterior Unruh radiation. To recover the standard
picture of gravitational collapse the two solutions should be superposed. However, Hawking's proposed resolution of the AMPS 
paradox is captured by the first of these solutions. 

If we take seriously the possibility that continued collapse doesn't occur, we expect to end up with a spherically symmetric, 
quasi-stable, static configuration of finite size. While no classical extended dust configuration can exist, we will argue that 
the interior Unruh radiation that accompanies collapsing dust shells will generate the conditions necessary for such a static 
configuration. The outgoing Unruh radiation leads to a negative mass singulaty, weakening the gravitational field, and may 
eventually cause the matter to expand again \cite{cvaz}.

\section{Dust collapse}

\subsection{Classical Solutions}

Dust collapse in 2+1-dimensions with a negative cosmological constant is described by the LTB family of solutions \cite{ltb}. In 
comoving and synchronous coordinates,$(\tau,\rho,\phi)$, the metric takes the form
\beq
 ds^2=-d\tau^2+\frac{(\partial_{\rho}R)^2}{2(E-F)}d\rho^2+R^2d\phi^2
 \label{LTB}
\eeq
where the physical radius of shells is given by $R(\tau,\rho)$ which obeys 
\beq
 (\partial_{\tau}R)^2=2E-\Lambda R^2.
 \label{1}
\eeq
The energy density, $\varepsilon(\tau,\rho)$, is given by
\beq
 2\pi G\varepsilon(\tau,\rho)=\frac{\partial_{\rho}F}{R(\partial_{\rho}R)},
 \label{2}
\eeq
where $\Lambda$ is the cosmological constant and $G$ is Newton's constant in $2+1$-dimensions. Using the freedom to rescale 
the shell labels, $\rho$, we can set $R(0,\rho)=\rho$ at some initial epoch. In this case the functions $F(\rho)$ and $E(\rho)$ 
may be given as
\bea
 F(\rho)&=2\pi G\int_0^{\rho}\rho'\varepsilon(0,\rho')d\rho'\nonumber\\
 E(\rho)&=[\partial_{\tau}R(0,\rho)]^2+\Lambda \rho^2
 \label{3}
\eea
The physical interpretation of these relations is that $2F(\rho)$ is the gravitational mass inside the shell labeled by $\rho$ 
and $E(\rho)/2$ is total energy per unit mass of that shell. Owing to this they are called the ``mass function'' and the 
``energy function'' respectively. We assume that $F(\rho)$ is a positive, monotonic increasing function of $\rho$ and 
that the initial data disallow shell crossings, {\it i.e.,} $R'(\tau,\rho) > 0$.

The solution to \eqref{1} that represents a collapsing dust ball is given by
\beq
R(\tau,\rho) = \sqrt{\frac{2E}\Lambda} \sin\left(-\sqrt{\Lambda}\tau + \sin^{-1} \sqrt{\frac{\Lambda\rho^2}{2E}}\right)
\eeq
and shows that the collapse inevitably forms a central singularity as each shell shrinks to zero physical radius at the proper 
time 
\beq
\tau_0(\rho)= \frac 1{\sqrt{\Lambda}}\sin^{-1}\left(\sqrt{\frac{\Lambda\rho^2}{2E}}\right).
\eeq
A detailed analysis \cite{ltb} also shows each shell, labeled by $\rho$, becomes trapped when its physical radius crosses the 
apparent horizon at $\Lambda R^2 -2F = 0$, {\it i.e.,} when $R<\sqrt{2F/\Lambda}$. Thus only shells satisfying the condition $F>0$ 
( therefore $\rho>0$) will become trapped, each at proper time
\beq
\tau_\text{ah}(\rho) = \frac 1{\sqrt{\Lambda}} \left(\sin^{-1} \sqrt{\frac{\Lambda \rho^2}{2E}} - \sin^{-1} \sqrt{\frac FE}\right).
\eeq
Moreover, by our assumptions about $F$, the physical radius of the apparent horizon will be a monotonic increasing function of 
$\rho$. Notice that $\tau_\text{ah}(\rho)<\tau_0(\rho)$, so each trapping surface forms before the shell becomes singular and
collapse to the central singularity is not a necessary condition for the formation of a trapping surface\footnote{To allow for 
an initial velocity profile that vanishes at the origin, the mass function is sometimes given as 
\beqn
F(\rho) =2\pi G\int_0^{\rho}\rho'\varepsilon(0,\rho')d\rho' - f_0,
\eeq
where $f_0$ is a positive integration constant. In this case, there will be a critical shell for which $F(\rho_c)=0$. It can then be shown 
that the singularity is timelike for $\rho<\rho_c$, null for $\rho=\rho_c$ and spacelike for $\rho>\rho_c$. In 2+1 dimensions, 
this does not lead to singular initial data and $f_0$ does not have the interpretation of a point mass at the center \cite{ltb}.}

\subsection{Collapse Wavefunctionals}

The canonical dynamics of collapsing dust shells is described by embedding the spherically symmetric ADM metric 
\beq
 ds^2=-N^2dt^2+L^2(dr+N^rdt)^2+R^2d\phi^2
 \label{4}
\eeq
in the LTB metric \eqref{LTB}. Above,  $N$ is the lapse, $N^r$ is the shift. After a series of canonical transformations \cite{}, 
they 
are described in a phase space of dust proper time, $\tau(t,r)$, the physical shell radius, $R(t,r)$, the mass dendity, 
$\varGamma(r)=F'(r)$, and their conjugate momenta,$P_{\tau}(t,r)$, $P_R(t,r)$ and $P_{\varGamma}(t,r)$ respectively. In this 
phase space the Hamiltonian and diffeomorphism constraints are\cite{vgks07}
\bea
 {P_{\tau}}^2+\mathcal{F}{P_R}^2-\frac{\varGamma^2}{\mathcal{F}}&=0\nonumber\\
 R'P_R-\varGamma P_{\varGamma}'+\tau'P_{\tau}&=0
 \label{5}
\eea
where the prime denotes a derivative w.r.t. the ADM radial coordinate, $r$, and
\beq
 \mathcal{F}\equiv\Lambda R^2-2F
 \label{6}
\eeq
The apparent horizon occurs when $\mathcal{F}=0$, which is determined by the vanishing of the null divergence. On the 
apparent horizon the physical radius of each shell is 
\beq
 R(\tau_{ah},\rho)=\sqrt{\frac{2F}{\Lambda}}
 \label{7}
\eeq
To transform from classical to quantum, Dirac's quantization procedure may be applied on the constraints, which act on wave 
functionals. The Hamiltonian constraint gives the Wheeler-DeWitt equation and the momentum constraint imposes diffeomorphism 
invariance. We begin with an ansatz for the wave functional
\beq
\Psi[\tau,R,\varGamma]=\exp\left[-i\int dr\varGamma(r)\mathcal{W}(\tau(r),R(r),\varGamma(r))\right]
\label{8}
\eeq
which automatically obeys the momentum constraint provided $\mathcal{W}$ doesn't have any explicit dependence on $r$. At 
this point we will consider a one-dimensional lattice with discrete points $r_i$ , a distance $\sigma$ apart. With this 
discretization and the ansatz in \eqref{8} the WDW equation gives
\beq
 \omega_i^2\left[\left(\frac{\partial\mathcal{W}_i}{\partial\tau_i}\right)^2+\mathcal{F}_i\left(\frac{\partial\mathcal{W}_i}
 {\partial R_i}\right)^2+\frac{1}{\mathcal{F}_i}\right]+\omega_i\left[\frac{\partial^2\mathcal{W}_i}{\partial\tau_i^2}+
 \mathcal{F}_i\frac{\partial^2\mathcal{W}_i}{\partial R_i^2}+A_i\frac{\partial\mathcal{W}_i}{\partial R_i}\right]+B_i=0
\eeq
Defining $\mathcal{W}_i=-iW_i$ and equating each co-efficient to zero as it is true for arbitrary $\omega_i$, we have three 
independent equations
\bea
 \left(\frac{\partial W_i(\tau_i,R_i,\varGamma_i)}{\partial\tau_i}\right)^2+\mathcal{F}_i\left(\frac{\partial W_i(\tau_i,R_i,
 \varGamma_i)} {\partial R_i}\right)^2=\frac{1}{\mathcal{F}_i}\nonumber\\
 \frac{\partial^2W_i(\tau_i,R_i,\varGamma_i)}{\partial\tau_i^2}+\mathcal{F}_i\frac{\partial^2W_i(\tau_i,R_i,\varGamma_i)}
 {\partial R_i^2}+A_i (R_i,\varGamma_i)\frac{\partial W_i(\tau_i,R_i,\varGamma_i)}{\partial R_i}=0 
  \label{15}\\
  B_i(R_i,\varGamma_i)=0\nonumber
\eea
whose solution is
\beq
 W_i=a_i\tau_i+\int dR_i\frac{\sqrt{1-a_i^2\mathcal{F}_i}}{\mathcal{F}_i}
\eeq
where $a_i=1/\sqrt{2(E_i-F_i)}$. Therefore, we have 
\beq
 \Psi=\lim_{\sigma\to 0}\prod_{i}\Psi_i(\tau_i,R_i,\varGamma_i)=\lim_{\sigma\to 0}\prod_{i}e^{\omega_ib_i}\times\exp
 \left\{-i\omega_i\left[a_i\tau_i\pm\int^{R_i}dR_i\frac{\sqrt{1-a_i^2\mathcal{F}_i}}{\mathcal{F}_i}\right] \right\}
 \label{9}
\eeq
with a well-defined continuum limit, where $\omega_i=\sigma\varGamma_i$.

The solutions are defined everywhere except at the apparent horizon where there is an essential singularity. Thus there are 
``exterior'' wave functionals that must be matched to ``interior'' wave functionals at the apparent horizon. A standard technique 
is to analytically continue the solutions into the complex plane. This technique is used to derive Hawking radiation as a 
tunneling process \cite{parikh}. Analytically continuing to the complex $R$-plane, following a semicircular contour in the 
upper-half plane of radius approaching zero around the pole, we find two sets of matched solutions,
\beq
  \Psi_i^{(1)} = \left\{ 
  \begin{array}{l l}
    e^{\omega_ib_i}\times\exp\left\{-i\omega_i\left[a_i\tau_i+\int^{R_i}dR_i\frac{\sqrt{1-a_i^2\mathcal{F}_i}}{\mathcal{F}_i}
    \right] \right\} & \quad \text{$\mathcal{F}_i>0$}\\
    e^{-\frac{\pi\omega_i}{g_{i,h}}}\times e^{\omega_ib_i}\times\exp\left\{-i\omega_i\left[a_i\tau_i+\int^{R_i}dR_i\frac{
    \sqrt{1-a_i^2\mathcal{F}_i}}{\mathcal{F}_i}\right] \right\} & \quad \text{$\mathcal{F}_i<0$}
  \end{array} \right.
  \label{10}
\eeq
and
\beq
  \Psi_i^{(2)} = \left\{ 
  \begin{array}{l l}
    e^{-\frac{\pi\omega_i}{g_{i,h}}}\times e^{\omega_ib_i}\times\exp\left\{-i\omega_i\left[a_i\tau_i-\int^{R_i}dR_i\frac{
    \sqrt{1-a_i^2\mathcal{F}_i}}{\mathcal{F}_i}\right] \right\} & \quad \text{$\mathcal{F}_i>0$}\\
    e^{\omega_ib_i}\times\exp\left\{-i\omega_i\left[a_i\tau_i-\int^{R_i}dR_i\frac{\sqrt{1-a_i^2\mathcal{F}_i}}{\mathcal{F}_i}
    \right] \right\} & \quad \text{$\mathcal{F}_i<0$}
  \end{array} \right.
  \label{11}
\eeq
where 
\beq
g_{i,h}=\partial_R\mathcal{F}_i(R_{i,h})/2
\label{11a}
\eeq
is the surface gravity of the $i$-th shell at the apparent horizon. 
The first set $\eqref{10}$ represents a flow toward the apparent horizon on both sides of it, that is, an infalling shell in 
the exterior is accompanied by an interior outgoing shell with relative probabilty determined by the Boltzman factor at 
the Hawking temperature of the shell. The second set $\eqref{11}$ represents a flow away from the apparent horizon on both 
sides, that is, an infalling interior shell representing continued collapse past the apparent horizon to the central 
singularity, is accompanied by an exterior outgoing shell with relative probabilty determined by the same factor. 
It represents thermal (Unruh) radiation in the exterior. Some useful conclusions can be drawn from the solutions. As mentioned 
in the introduction, one may superpose both the solutions leading to a picture in which continued collapse to a 
singularity occurs, with accompanying thermal radiation in the exterior. However if we take only $\eqref{10}$ as the basis 
for quantum collapse then there will be thermal Unruh radiation inside the apparent horizon but no thermal radiation outside. 
There will be no continued collapse to a singularity. The collapse would terminate at the apparent horizon $(\mathcal{F}_i=0)$, 
which agrees with Hawking's proposal.

\subsection{A quasi-classical configuration}

As we see from \eqref{10} collapse stops at the apparent horizon forming a quasi-stable compact object, we expect that there exist 
solutions to the Einstein equations 
with finite boundary radius which show the effect of the internal Unruh radiation. The matter should condense on the 
apparent horizon and the solution should match smoothly with the external BTZ vacuum. Within the dust ball, the metric will have 
the form,
\beq
 ds^2=-e^{2A(r)}dt^2+e^{2B(r)}dr^2+r^2d\phi^2\nonumber\\
 \label{12}
\eeq
and the corresponding field equations, with ${T^{\mu}}_{\nu}=\diag\{-\varepsilon(r),p_r(r),p_{\phi}\}$, will be
\bea
&\frac{e^{2(A-B)}B'}{r}+\Lambda e^{2A}=4\pi Ge^{2A}\varepsilon(r)\nonumber\\
&\frac{A'}{r}-\Lambda e^{2B}=4\pi Gp_r(r)\nonumber\\
&e^{-2B}r^2(A''-A'B'+A'^2)-\Lambda r^2=4\pi Gp_{\phi}(r)
\label{13}
\eea
where $\varepsilon(r)$  is the energy density within the dust ball, $p_r(r)$ is the radial pressure and $p_{\phi}(r)$ is the 
tangential pressure. We can choose two of the stress-enegy tensor components arbitrarily and the third will be determined 
by the Einstein equations. We choose the energy density and set the tangential pressure to zero. 
According to equation $\eqref{6}$, the mass function that is expected of a dust ball whose collapse has stopped at the apparent 
horizon is $F(r)=\frac{\Lambda r^2}{2}$. This shows that the energy density will be
\beq
 \varepsilon(r)=\frac{\Lambda}{2\pi G},
 \label{16}
\eeq
so the $tt$-component of the field equations gives
\beq
 e^{2B}=\frac{1}{C_1-\Lambda r^2}.
 \label{17}
\eeq
For a physically meaningful solution, $C_1$ has to be greater than $\Lambda r^2$. We will see later that it actually 
describes a negative point mass source at the center. With this solution for $B(r)$ if we solve the tangential component 
of the field equations we find
\beq
 e^{2A}=\cosh^2\left(\arctan\left(\frac{\sqrt{\Lambda}r}{\sqrt{C_1-\Lambda r^2}}\right)+\tilde{C_2}\right)
 \label{18}
\eeq
A singularity occurs at $r=0$. We can calculate the radial pressure directly from the $rr$-component of the field equations
\beq
p_r=\frac{\Lambda}{\Lambda r^2-C_1}+\frac{\sqrt{\Lambda}\tanh\left[\tilde{C_2}+\arctan\left(\frac{\sqrt{\Lambda}r}
{\sqrt{C_1-\Lambda r^2}}\right)\right]}{r\sqrt{C_1-\Lambda r^2}}.
\label{20}
\eeq
If $r_b$ denotes the outer boundary of the collapsed star, we want to match the interior geometry to the outer vacuum described by the BTZ metric
\beq
ds^2=-f(R)dT^2+\frac{1}{f(R)}dR^2+R^2d\phi^2
\label{21}
\eeq
where $f(R)=(\Lambda R^2-GM_s)$ and $M_s$ is the BTZ mass of the dust ball. The junction conditions require that, at $r_b=R_b$,
\beq
 dT=\sqrt{\frac{e^{2A(r_b)}}{f(r_b)}}dt\hspace{1cm}e^{-B(r_b)}=\sqrt{f(R_b)}\hspace{1cm}2A'(r_b)=(\ln f)'\mid_{R_b}
 \label{22}
\eeq
These give,
\bea
 &C_1 =\Lambda(2r_b^2-r_s^2)\nonumber\\
 &\tilde{C_2}=-\tan^{-1}\left(\frac{r_b}{\sqrt{r_b^2-r_s^2}}\right)+\tanh^{-1}\left(\frac{r_b}{\sqrt{r_b^2-r_s^2}}\right)
 \label{23}
\eea
where we define $GM_s=\Lambda r_s^2$.

\section{Energy extraction}

Our solutions depend on two constants, which can be taken to be the BTZ radius, $r_s$, and the boundary 
radius, $r_b$, of the dustball. The two are related by the strength of the negative mass singularity 
at the center as, according to \eqref{23}, 
\beq
M_s=M_b-(C_1/G-M_b)
\label{23a}
\eeq
where the quantity $C_1/G-M_b=M_0$ is the mass energy extracted from the center of the dustball by the outgoing Unruh radiation.
To estimate the value of $M_0$, we assume each shell to be a quantum harmonic oscillator with definite energy. A 
single quantum harmonic oscillator, located at lattice point $i$, will have mean energy
\beq
 \langle E_i\rangle=\frac{\omega_i}{2}\coth(\beta_i\omega_i/2)
\eeq
where $\beta_i=\frac{1}{kT_i} = \frac 1{\Lambda r_i}$. Moreover, for our collapse,
\beq
 \omega_i=\frac{\sigma\Lambda r_i}{G}
\eeq
and therefore $\beta_i\omega_i=\frac{\sigma}{G}$. At this point we should notice that $\sigma$ cannot be arbitrarily 
small in the continuum limit as the total energy, $E=\sum_i \langle E_i\rangle$, would then be unbounded. Instead, it will 
be such that it is microscopically 
large but macroscopically small so that a given lattice spacing contains many Planck lengths. Therefore, assuming 
$\sigma\gg G$, the average energy of the Unruh radiation inside the apparent horizon will be ($r_i = i \sigma$)
\beq
 \langle E\rangle \approx \frac 12\sum_{i=1}^N \omega_i=\sum_{i=1}^N\frac{\Lambda i \sigma^2}{2G} = \frac{\Lambda N(N+1) 
 \sigma^2} {4G} = \frac{\Lambda r_b^2}{4G}
 \label{25}
\eeq
where $N$ is the number of shells to the boundary, {\it i.e.,}  $r_b = N\sigma$. Equating the mean energy of the 
Unruh radiation to the mass $M_0$ in $\eqref{23a}$, we find that $C_1=5\Lambda r_b^2/4$ 
and therefore 
\beq
r_s=\frac{\sqrt{3}r_b}{2},
\eeq
so the BTZ (horizon) radius lies within the boundary of the star and the collapse does not end up forming an event 
horizon, which supports Hawking's conjecture. 

\section{Conclusion}

In this paper, we have speculated about what may be expected from quantum collapse in 2+1 dimensions with a 
cosmological constant. Our interest in this model stems from the fact that the 2+1 dimensional model has served in 
the past as a toy model of black hole thermodynamics and as an aid to understanding many of the higher dimensional 
black holes of string theory. Long ago, Strominger \cite{str98} argued 
that because the asymptotic symmetry group of 2+1 dimensional gravity with a cosmological constant is generated by 
two copies of the Virasoro algebra, its degrees of freedom can be described by a two dimensional conformal field theory 
(CFT) at infinity, with central charges $c_R=c_L = 3l/2G$. Ever since then, most approaches to describing the BTZ black 
hole have employed the AdS/CFT correspondence. A connection between the description of the quantum BTZ black hole via 
the canonical approach and its description via the AdS/CFT correspondence was later found \cite{vgksw08}.

Here we have shown, however, that the collapse process need not lead to the formation of a black hole. It is 
possible, within the quantum description, that collapse stops when all shells arrive at the apparent horizon.
The quantum wavefunctionals indicate that, during the collapse, infalling shells of matter never cross the apparent 
horizon and are accompanied by outgoing Unruh radiation emanating from the center, which also terminates at the apparent 
horizon. There is no exterior Unruh radiation. The effect of this outgoing radiation and the absence of continued 
collapse across the apparent horizon leaves behind a negative mass singularity at the center, as the system settles 
into a quasi-stable, static configuration. We have found static solutions smoothly matching the BTZ vacuum at the boundary
and possessing an energy density describing dust which has coalesced on the apparent horizon together with radial 
pressure, which is presumed to originate in quantum gravity.  These solutions are fully determined by two parameters, 
{\it viz.,} the radius of the dust ball and its BTZ radius, which differ by the magnitude of the mass of the central 
singularity. The boundary lies outside the BTZ (horizon) radius and no Unruh radiation escapes, therefore the information 
loss problem gets resolved.

The next step is to determine a dynamical collapse model whose end state is the configuration described in this article.
Ideally, we would like to find an effective collapse scenario that incorporates the features predicted by the wavefunctional 
in \eqref{10}, achieves the quasi-stable configuration described in this article and eventually expands, thus preserving CPT
invariance. It would be worthwhile finding such a description within the context of the AdS/CFT correspondence.


\begin{thebibliography}{99}
\bibitem{bek72}J.D. Bekenstein, Ph.D. thesis, Princeton University (1972);\\
{\it ibid} Lett. Nuovo Cimento {\bf 4} (1972) 373;\\
{\it ibid} Phys. Rev. D {\bf 7} (1973) 2333.
\bibitem{haw75}S.W. Hawking Comm. Math. Phys. {\bf 43}(1975) 199
\bibitem{sus}L. Susskind, L. Thorlacius, J. Uglum, Phys. Rev. {\bf D} 48 (1993) 3743
\bibitem{thex}D-h Yeom, H. Zoe, Phys. Rev. {\bf D} 78 (2008) 104008; {\it ibid} Int. J. Mod. Phys. {\bf A} 26 (2011) 3287\\
S. Braunstein et al. Phys. Rev. Lett. {\bf 110} (2013) 101301
\bibitem{amps12}A. Almehri, D. Marolf, J. Polchinsky, J. Sully, [arXiv:1207.3123]
\bibitem{haw12}S. W. Hawking, [arXiv:1401.5761]
\bibitem{ltb}Sashideep Gutti,Class. Quant. Grav. {\bf 22} (2005) 3223.
\bibitem{wdw}C. Vaz, L. C. R. Wijewardhana, Phys. Rev. D {\bf 82} (2010) 084018;\\ C. Vaz, Int. J. Mod. Phys. D {\bf 23} 
(2014) 1441002.
\bibitem{cvaz}C. Vaz, Nucl. Phys. {\bf B} 891 (2015) 558
\bibitem{vgks07} C. Vaz, S. Gutti, C. Kiefer and T.P. Singh, Phys. Rev. D {\bf 76} (2007) 124021. 
\bibitem{parikh}Maulik K. Parikh and Frank Wilczek, Phys. Rev. Lett. {\bf 85} (2000) 5042-5045 
\bibitem{lcr} C. Vaz and L.C.R. Wijewardhana, Class. Quant. Grav. {\bf 27} (2010) 055009 
\bibitem{str98}A. Strominger, JHEP 9802 (1998) 009.
\bibitem{vgksw08}C. Vaz, S. Gutti, C. Kiefer, T.P. Singh and L.C.R. Wijewardhana, Phys. Rev. D {\bf 77} (2008) 064021. 
\end{thebibliography}
\end{document}